\documentstyle[12pt,epsfig]{article}
\begin{document}
\begin{center}
{\large \bf 
Stringent New Bounds on Supersymmetric Higgs Bosons from
Existing Tevatron Data} \\
\vspace*{6mm}
Manuel Drees$^1$, Monoranjan Guchait$^{2,3}$, and Probir Roy$^3$ \\
$^1${\it APCTP, 207--43 Cheongryangryi--dong, Tongdaemun--gu, Seoul
130--012, Korea} \\
$^2${\it Physics Department, Yonsei University, Seoul 120--749, Korea} \\
$^3${\it Tata Institute of Fundamental Research, Mumbai, India} \\
\end{center}
\vspace*{1cm}

There is an error in our reported calculation \cite{old} of the cross
section for associated $b \bar{b}$~Higgs production in hadron
colliders. After correction, the cross section at the Tevatron is
reduced by a factor of about 10. (Our calculations have been performed
by choosing the Higgs mass as the QCD scale and CTEQ3L structure
functions; there could be an uncertainty of upto 40\% from different
choices). This weakens (increases) the upper bound on $\tan \! \beta$
that can be derived from the existing CDF data \cite{cdf} by slightly
more than a factor of 3. The excluded region, in the plane spanned by
the MSSM parameters $m_A$ and $\tan \!  \beta$, is shown in Fig.~1,
which corrects Fig.~2 of our original publication \cite{old}. We are
still able to extend the region that can be excluded from a
consideration of $t \rightarrow H^+$ decays \cite{dp}, though our
upper bound on $\tan \! \beta$ is always above 80. However, a similar
analysis, applied to the forthcoming data on $b \bar{b} \tau^+
\bar{\tau^-}$ final state events from RUN 2 at the Tevatron, would be
able to lead to much stronger conclusions.

We thank Steve Mrenna for alerting us to the problem in our
calculation, and Michael Spira for providing us with results of his
independent calculation.

\section*{Figure Captions}
{\bf Fig.~1:} Constraints on the MSSM Higgs sector in the
$(m_A, \ \tan \! \beta)$ plane. The region $m_A < 75$ GeV is excluded
by Higgs searches at the LEP collider. The region above the dashed curve
is excluded by an analysis [2] of top quark decays. The region
above the solid line is excluded by our analysis using CDF limits on
$b \bar{b} \tau^+ \tau^-$ final states. For $\tan \! \beta \gg 1$, the
CP--odd scalar $A$ is nearly degenerate with either the light CP--even
scalar $h$ or the heavy CP--even state $H$; the cross--over occurs between
100 and 125 GeV, depending on the details of the sparticle spectrum.

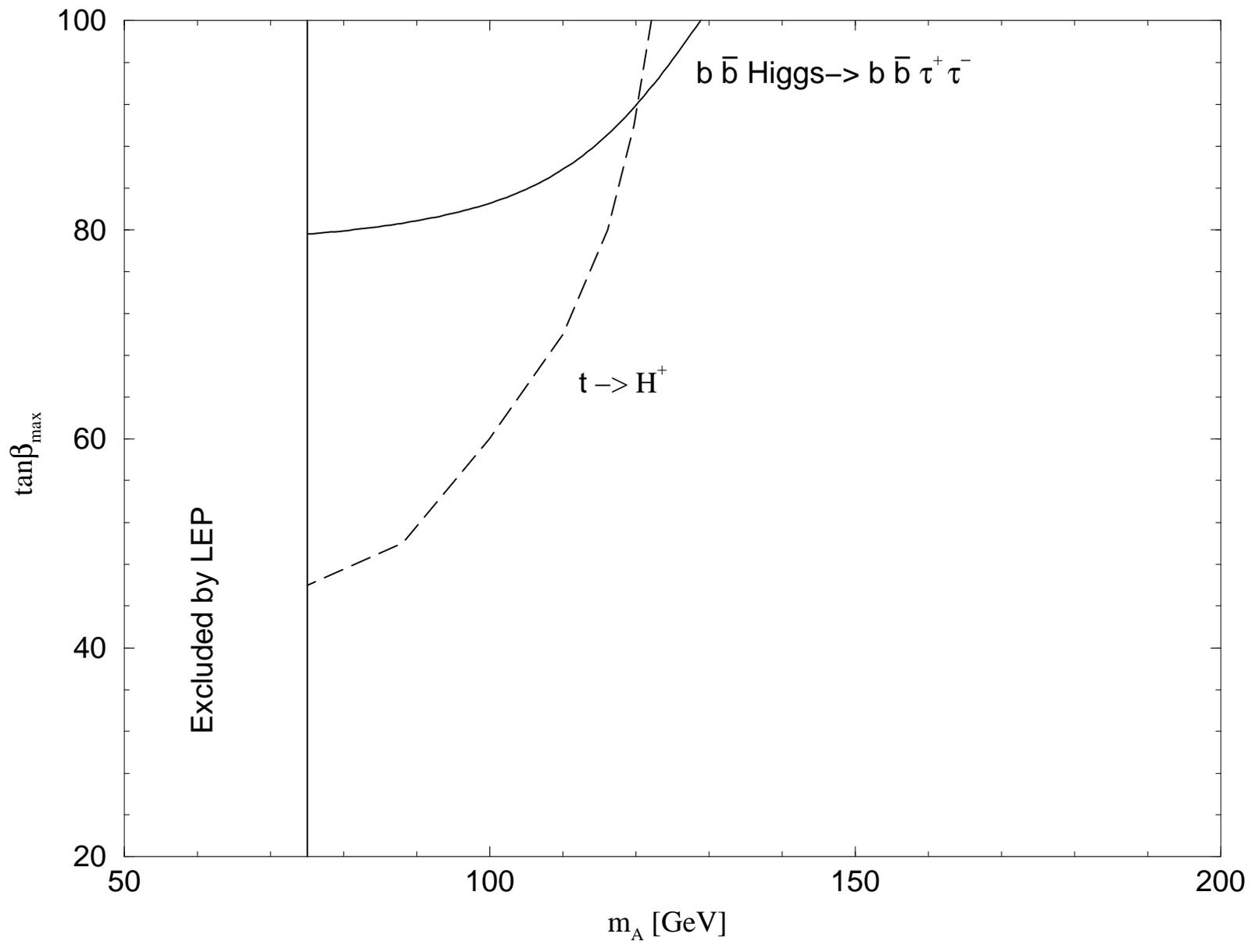
\end{document}